# Experimental Two Way Quantum Key Distribution Protocol with Decoy State

**M. F. Abdul Khir[1,3], M. N. Mohd Zain[3], Iskandar Bahari[4], Suryadi[2], S. Shaari[1]**

[1] Photonic Lab, IMEN, Universiti Kebangsaan Malaysia, 43400 UKM Bangi, Malaysia
[2] Faculty of Science, International Islamic University of Malaysia (IIUM), Jalan Bukit Istana, 25200 Kuantan, Pahang, Malaysia
{[3] Photonic Lab, Micro Systems and MEMS, [4] Advance Analysis and Modelling}, MIMOS Berhad, Technology Park Malaysia,
57000 Kuala Lumpur, Malaysia

**ABSTRACT**

We report on the first demonstration of a two way Quantum Key Distribution protocol with decoy state. The experiment was conducted over free space medium and exhibits a significant increase in the maximum secure distance.

## 1. INTRODUCTION

Since the introduction of BB84 protocol in 1984 [1], Quantum Key Distribution (QKD) combined with one time pad presents a likely candidate for an unconditionally secure method for information transfer between two distant parties. Over the last decade, practical aspect has been at the center of interest in Quantum Key Distribution (QKD) where much of the efforts are in bridging the gap between the imperfect settings of QKD realization to the theoretical unconditional security proof [2,3]. While the imperfect settings is known to invite powerful attacks such as Photon Number Splitting (PNS) attack, it does not necessarily form a threat to the security of the shared key but rather limits the secure key generation and the maximum secure distance.

One recently introduced tool that equipped QKD against such sophisticated attacks is the decoy state method, first introduced by [4] and further developed for example by the work in [5-11]. In decoy state method, besides signal pulses, Alice and Bob use several other different intensities of coherent light pulses as decoy states. Since Eve cannot distinguish between the signal and decoy pulses, she has to apply the same strategy to all of them. As a result, any eavesdropping attempt by Eve will inevitably modify the photon statistic and expose her. [5].

Theoretical as well as experimental progresses with regard to decoy state can be seen for example by works in [5-18]. One common aspect in these works is that they were all based on BB84 protocol. As for other protocol in prepare and measure scheme, extension to decoy state can be seen in [12,13] for SARG04 protocol. In the case of two way QKD protocols [19-27], recently, Shaari et al in [14] started the decoy state extension in particular with the LM05 which is a four state two way protocol by Lucamarini and Mancini [22,24]. They proposed a two decoy states extension as a practical decoy scheme for LM05. The result of their numerical analysis was quite encouraging given that the maximum secure distance is extended by almost double compared to the one without decoy state.

In this work, we use the bounds obtained from our work in [31] in which we have extended the work in [14] to the case of "weak+vacuum", first proposed by Lo et al for BB84 in [7]. While it has been shown in [7] that this special case of practical decoy states is in fact optimal for BB84, experimentally, the proposed scheme has obvious advantages in simplifying the setup at state preparation stage due to for example reduce number of photon source. While one may encounters problem with producing a "truly vacuum" state in a plug and play QKD setup due to difficulties such as in finding good attenuator, it is not the case for a one way QKD setup as it can be easily achieved by simply not triggering the laser sources which in this case fits well to our experimental setup. We then experimentally demonstrate an implementation of LM05 protocol with the proposed weak+vacuum decoy state over free space medium. To the best of our knowledge this is the first time that an implementation of a two way QKD protocol with decoy state is realized. This letter is organized as follows. The ensuing sections introduce the proposed protocol as well as elaborations of the experimental setup. We then present our main results and analysis and conclude with suggestions for future works.

## 2. THE PROTOCOL

In [31], we have simplified relevant bounds that led to the two secure key rate formulas represented by Eq 25 and Eq 26 in [14] and obtain for the case of "weak+vacuum". While the former results in higher secure key rate and longer secure distance, the latter has an advantage of not having to concern on how Eve may manipulate the single and double photon contributions individually [14]. Note also that the former requires additional information by having to rely on the value of double photon yield ($Y_2$) from infinite case denoted as $Y_2^\infty$ in Eq.10 of [14] to lower bound $Y_2^L$. As such, for simplicity, we make use of the latter throughout our experiment.

Let us review the relevant bounds used in obtaining the lower bound of key generation rate $R$. The lower bound $(Y1 + Y2)^L$ for the case of "weak+vacuum" adapted from Eq. 19 of [14], is given by :

$$(Y1+Y2)^L = \frac{\mu^3 e^\nu Q_\nu - \nu^3 Q_\mu e^\mu - (\mu^3 - \nu^3)Y_0 + \left(\nu^3\mu - \frac{1}{2}\nu^3\mu^2\right)Y_1^L}{\mu^3(\nu - \frac{1}{2}\frac{\nu^3}{\mu})} \tag{1}$$

where $Y_1^L$ is given by [7] as :

$$Y_1 \geq Y_1^L = \frac{\mu}{\mu\nu - \nu^2}\left(Q_\nu e^\nu - Q_\mu e^\mu \frac{\nu^2}{\mu^2} - \frac{\mu^2 - \nu^2}{\mu^2}Y_0\right) \tag{2}$$

The lower bound of effective gain ($Q_{12}^L(\mu)$) and upper bound of effective error rate ($\varepsilon^U$) is given by [14] as :

$$Q_{12}^L(\mu) = \left[\frac{(Y1+Y2)^L}{2}\mu^2 + (Y_1^L\mu - \frac{Y_1^L\mu^2}{2})\right]e^{-\mu} \tag{3}$$

$$\varepsilon^U = \frac{E_\mu Q_\mu - e_0 Y_0 e^{-\mu}}{Q_{12}^L} \tag{4}$$

The effective gain ($Q_{12}^L(\mu)$) and error rate ($\varepsilon^U$) can be plugged into the following Eq 5 for the lower bound of key generation rate, given by [14] as :

$$R \geq R^L = -Q_\mu f(E_\mu) H(E_\mu) + Q_{12}^L[1 - \tau(\varepsilon^U)] \qquad (5)$$

where

$H(E_\mu)$ is the binary Shannon Entrophy and is given by

$$H(E_\mu) = -E_\mu \log_2(E_\mu) - (1 - E_\mu)\log_2(1 - E_\mu) \qquad (6)$$

and $\tau(e)$ as

$$\tau(e_1) = \begin{cases} \log_2(1 + 4e_1 - 4e_1{}^2), & e_1 < \frac{1}{2} \\ 1 & , \quad e_1 \geq \frac{1}{2} \end{cases} \qquad (7)$$

### 3. EXPERIMENTAL SETUP

The schematic of our experimental setup is depicted in Fig. 2. It is actually an extension of our previous setup for an automated LM05 protocol implementation over free space medium in [29]. We made several modifications at the source as well as at the LabVIEW program to accommodate the proposed decoy state extension.

Our LabVIEW 8.5 based program, developed on top of a 40 MHz Reconfigurable I/O module of National Instruments (PXI-7833R) pair at Bob and Alice controls and synchronizes all the laser sources, Pockels cells and single photon counting modules. The program used pseudo-random number generator to set random triggering of each laser source and Pockels cell. For the case of weak+vacuum decoy state, we set such that the pulses for weak decoy state, vacuum state and signal state are randomly distributed at 1:1:2. The pulse repetition rate of the FPGA was set to be 0.725 MHz so that they operate within the safe region of below 1 MHz limitation of our Pockels cells.

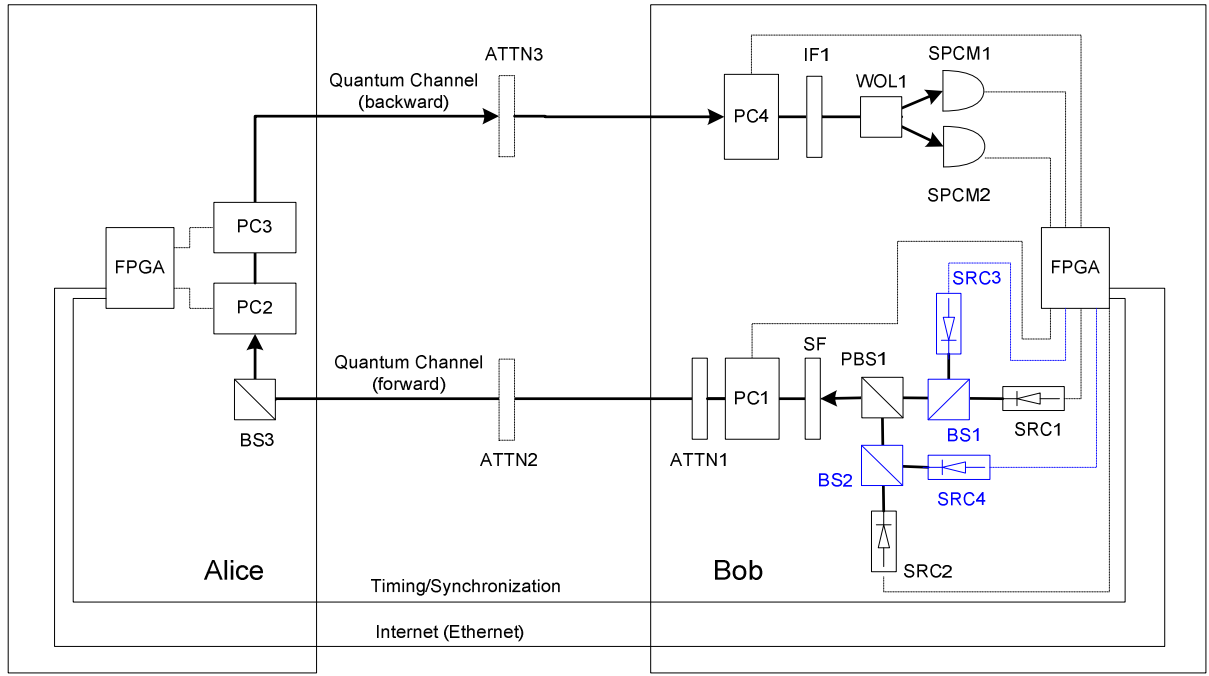

Fig. 2. The LM05 and decoy state experimental setup consists of SRC1, SRC2, photon source for signal state; SRC3,SRC4, photon source for decoy state ; BS1, BS2, BS3 beam splitter;PBS1, polarization beam splitter; SF, spatial filter; PC1, first Pockels cell; ATTN1, variable attenuator; BS1,BS2 50/50 beam splitter; ATTN2, ATTN3 attenuator; PC2, second Pockels cell; PC3, third Pockels cell; PC4, Fourth Pockels cell; IF1, interference filter; WOL1, Wollaston Prism; SPCM1, H & D detector; SPCM2, V & A detector.

The photon source consists of four diode lasers (SRC1, SRC2, SRC3, SRC4) from Coherent CUBE with wavelength at 785 nm. The SRC1 and SRC2 pair is assigned as the source for signal state which emits either horizontal or vertical pulse respectively while the SRC3 and SRC4 pair is assigned for the decoy source. They are all separately attenuated to achieve the required intensities. Either one of these four lasers is randomly triggered at a time. Horizontal pulses from SRC1 and SRC3 are combined into horizontal optical path at BS1 while vertical pulses from SRC3 and SRC4 are combined into vertical optical path at BS2. Optical pulses from the two paths are finally combined into a same optical path at polarization beam splitter PBS1 before being spatially filtered at SF. The optical pulses further proceed to Pockels cell PC1 whose function is to randomly transform incoming horizontal or vertical pulse to diagonal or anti-diagonal respectively. In this way, the combination of randomly triggered laser sources and PC1 produces the required signal and weak decoy states. For the case of vacuum state, Bob simply choose to not trigger any of the the laser as well as the pockels cells.

Each prepared photon pulse that reaches Alice will first go through the 50/50 beam splitter (BS3) used for simulating the effects of control mode and then proceed to the flipper (PC2,PC3) which is randomly flipped or not flipped by Alice for logical bit 1 or 0 encoding. The operation of the flipper is described in detail in [28,29]. Encoded pulses returning from Alice enter Bob's detection package by first going through the Pockels cell PC4 whose function is to actively prepare measurement basis to the one used at state preparation stage.. They are further polarization analyzed by the Wollaston prism WOL1 to be either detected at DET1 or DET2.

## 4. RESULTS AND DISCUSSION

It is known that with certain known intrinsic parameters i.e. internal transmission of the system including detection efficiency ($\eta_{Bob}$), erroneous detection probability ($e_{detector}$) and background rate ($Y_0$), one may perform numerical simulation to determine the optimal mean photon number for a particular distance or channel loss as well as estimating maximum secure distance capable with a QKD setup [7,15,16]. We notice that estimation of maximum secure distance is very important in this experiment as it is of no use to perform decoy state at a distance where secure key generation is not anymore possible. As such, we first obtained intrinsic parameters measured from the experimental setup in which we obtained $\eta_{Bob} = 0.072$, $e_{detector} = 0.045$ and $Y0 = 3.52 \times 10^{-6}$. We then performed numerical simulation to estimate the capability of the proposed decoy state extension with this setup in terms of maximum secure distance. Note that due to differences in the setup and alignment accuracy, they are different to those previously used in [29]. We have also numerically searched the maximum key generation rate at every distance for the case of without decoy state and theoretical infinite decoy state. For the case of without decoy state, we based on the key rate formula in [24] while for the case of theoretical infinite decoy state, we based on the key rate formula in [14,31]. The two formulas are denoted as $R_{LM}$ and $R_\infty$ respectively here.

In conducting the experiment, we started with the case of without decoy state followed by the case of "weak+vacuum". A same mean photon number closed to optimal as suggested by numerical simulation was used for every distance which is, specifically μ = 0.31 and ν = 0.13 for $R_{12}$ and $\mu = 0.15$ for $R_{LM}$. As for the distance, we have made use of the two attenuators denoted as ATTN2 and ATTN3 in Fig 1 to simulate the channel loss. The data size used for each experimental plot was 140 Mbit. The experimental result is shown in table 1.

Table 1. Experimental results for the case of "weak+vacuum" decoy state.

| Channel Loss (dB) | $Q_\mu$ | $E_\mu$ | $Q_\nu$ | $E_\nu$ | $Y_0$ |
|---|---|---|---|---|---|
| **1.24** | $1.182 \times 10^{-2}$ | $4.487 \times 10^{-2}$ | $5.245 \times 10^{-3}$ | $4.448 \times 10^{-2}$ | $4.383 \times 10^{-6}$ |
| **3.26** | $4.588 \times 10^{-3}$ | $4.141 \times 10^{-2}$ | $2.137 \times 10^{-3}$ | $4.180 \times 10^{-2}$ | $3.518 \times 10^{-6}$ |
| **5.23** | $2.059 \times 10^{-3}$ | $4.964 \times 10^{-2}$ | $8.470 \times 10^{-4}$ | $5.084 \times 10^{-2}$ | $3.251 \times 10^{-6}$ |
| **6.50** | $1.086 \times 10^{-3}$ | $5.397 \times 10^{-2}$ | $4.706 \times 10^{-4}$ | $5.410 \times 10^{-2}$ | $3.686 \times 10^{-6}$ |
| **8.38** | $4.995 \times 10^{-4}$ | $5.274 \times 10^{-2}$ | $2.120 \times 10^{-4}$ | $5.468 \times 10^{-2}$ | $3.918 \times 10^{-6}$ |
| **9.46** | $2.837 \times 10^{-4}$ | $6.215 \times 10^{-2}$ | $1.340 \times 10^{-5}$ | $6.850 \times 10^{-2}$ | $3.947 \times 10^{-6}$ |
| **11.01** | $1.303 \times 10^{-4}$ | $6.117 \times 10^{-2}$ | $5.711 \times 10^{-5}$ | $8.473 \times 10^{-2}$ | $4.005 \times 10^{-6}$ |

Using experimental results in Table 1, we then calculate the lower bounds of single and double photon gain($Q_{12}$), key rate ($R$) and upper bound of $e_{12}$ by plugging into Eq 1 ~ 7. The results are shown in Table 2. We used $f(E_\mu) = 1.22$ for error correction efficiency. To better illustrate the results of this experiment, we have also plot relevant graph depicted in Fig 2.

Table 2. The Lower bounds of $Q_{12}$, $R$ and upper bound of $e_{12}$. The values are calculated using Eq 1~7 with experimental data in Table 1.

| Channel Loss (dB) | $Q_{12}^L$ | $e_{12}^U$ | $R^L$ |
|---|---|---|---|
| 1.24 | $9.535 \times 10^{-3}$ | $5.546 \times 10^{-2}$ | $3.108 \times 10^{-3}$ |
| 3.26 | $3.518 \times 10^{-3}$ | $5.361 \times 10^{-2}$ | $1.188 \times 10^{-3}$ |
| 5.23 | $1.573 \times 10^{-3}$ | $6.421 \times 10^{-2}$ | $3.689 \times 10^{-4}$ |
| 6.50 | $8.310 \times 10^{-4}$ | $6.887 \times 10^{-2}$ | $1.560 \times 10^{-4}$ |
| 8.38 | $4.060 \times 10^{-4}$ | $6.135 \times 10^{-2}$ | $1.029 \times 10^{-4}$ |
| 9.46 | $2.340 \times 10^{-4}$ | $6.910 \times 10^{-2}$ | $4.058 \times 10^{-5}$ |
| 11.01 | $9.825 \times 10^{-5}$ | $6.625 \times 10^{-2}$ | $1.411 \times 10^{-5}$ |

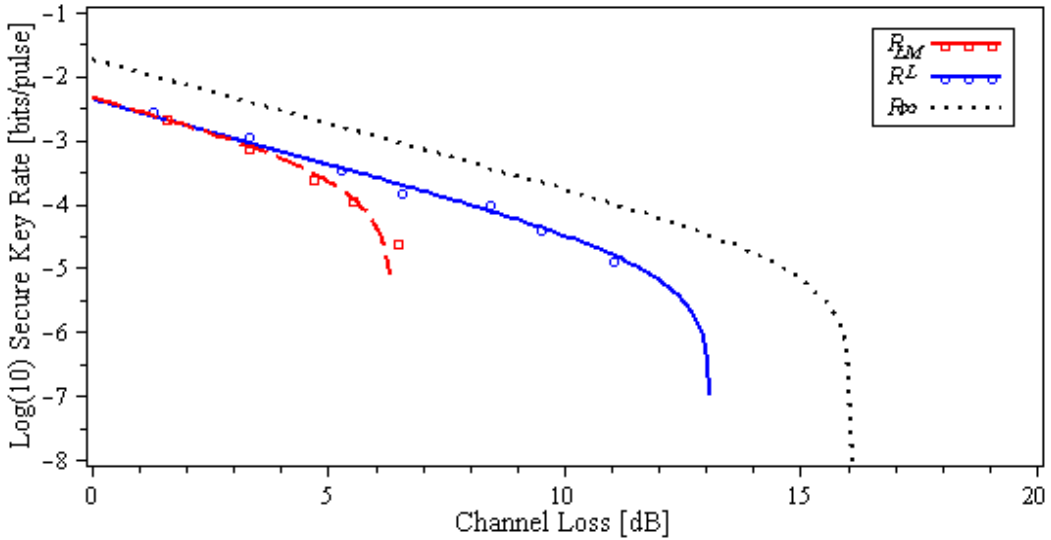

Fig. 2. Experimental plot for the case of $R_{LM}$, $R_{12}$ and $R_\infty$ using instrinsic parameters of the setup as in Table 1. The $R_{12}$ and $R_\infty$ are obtained using Eq 5 and $R_{LM}$ is obtained using key rate formula in [24] . The theoretical lines are based on numerical simulation results using intrinsic parameters $\eta_{Bob} = 0.072$ , $e_{detector} = 0.045$ and $Y_0 = 3.52 \times 10^{-6}$ and with mean photon number closed to optimal with $\mu = 0.31$ and $\nu = 0.13$ for $R_{12}$ and $\mu = 0.15$ for $R_{LM}$. For $R_\infty$, optimal µ was used throughout every distance.

From Fig 2, it is obvious that without decoy state, a maximum secure distance will only reach less than 8 dB channel loss. We verified at 9.51 dB that the secure key generation rate is negative for the case of without decoy state. In contrast, using the proposed extension of "weak+vacuum" decoy state, one could easily extend the maximum secure distance by around two third of the one without decoy state which fits well with our previous numerical simulation results in [31] when we used data from GYS [32].

## 5. CONCLUSIONS AND FUTURE WORKS

We have successfully implemented a QKD system based on a two way protocol namely the LM05 and decoy state method over a free space medium. The experimental result verified the capability of the proposed scheme which significantly extends the maximum secure distance by around two third of the one without decoy state. The fact that only a simple extension to existing setup was needed suggests the practicality of this method and deserves further study.

## ACKNOWLEDGEMENTS

The authors would like to thank Jesni Shamsul Shaari and Marco Lucamarini for fruitful discussion.